\documentclass[letter,twocolumn]{jpsj3}
\usepackage{bm,color}

\title{Disorder-Induced Multiple Transition involving 
$\mathbb{Z}_2$ Topological Insulator}

\author{Ai Yamakage\thanks{E-mail address: ai@cmpt.phys.tohoku.ac.jp}, Kentaro Nomura$^1$, Ken-Ichiro Imura$^2$, and Yoshio Kuramoto
}
\inst{Department of Physics, Tohoku University, Sendai 980-8578, Japan \\
$^{1}$Correlated Electron Research Group (CERG), RIKEN-ASI, Wako 351-0198, Japan \\
$^2$Department of Quantum Matter, AdSM, Hiroshima University, Higashi-Hiroshima 739-8530, Japan
}
\abst{Effects of disorder on two-dimensional $\mathbb{Z}_2$ topological insulator
are studied numerically by the transfer matrix method.
Based on the scaling analysis, the phase diagram is derived for a model of HgTe quantum well as a function of disorder strength and magnitude of the energy gap. 
In the presence of $s_z$ non-conserving spin-orbit coupling, a finite metallic region is found that partitions the two topologically distinct insulating phases.
As disorder increases,
a narrow-gap topologically trivial insulator 
undergoes a series of transitions; 
first to metal, second to topological insulator, third to metal, and finally back to trivial insulator.
We show that this multiple transition is a consequence of 
two disorder effects;
renormalization of the band gap,
and Anderson localization.
The metallic region found in the scaling analysis
corresponds roughly to the region of finite
density of states at the Fermi level evaluated in the
self-consistent Born approximation.}
\kword{$\mathbb Z_2$ topological insulator, HgTe, disorder, phase diagram, density of states, transfer matrix, self-consistent Born approximation}

\begin{document}
\maketitle

Topological insulator is a distinct type of insulator, showing a metallic surface state. 
During the last few years an increasing attention has been on its $\mathbb{Z}_2$ version
\cite{Z2TI, hasan10, qi10}
which realizes without an external magnetic field and preserves time-reversal symmetry. 
This is contrasting to a more prototypical integer quantum Hall insulator. 
Another twist motivating the present study is a recent proposal on the concept of 
so-called ``topological Anderson insulatorh (TAI).
\cite{TAI, jiang09, TTAI}
which extends the limit of topological distinction 
beyond that of band insulator.

In two spatial dimensions (2D) disorder plays a fundamental role in electronic transport. 
This paper deals with a 2D version of the $\mathbb Z_2$ topological insulator ($\mathbb{Z}_2$TI), 
a variant of the model proposed by
Bernevig, Hughes, and Zhang; hereafter referred to as BHZ.
\cite{BHZ}
The BHZ model has been proposed to describe a $\mathbb{Z}_2$TI
implemented in HgTe/CdTe superstructure.
\cite{HgTe,  HgTe_JPSJ}
An important element added in this study to the BHZ model 
is a Rashba-type spin-orbit coupling (SOC),
which breaks 
spin-axial symmetry along $z$-axis inherent to the original version 
of BHZ model.
This $s_z$ non-conserving SOC changes the topological invariant characterizing insulator from 
$\mathbb{Z}$ to $\mathbb{Z}_2$.
\cite{KM_Z2,KM_QSH}
% while upgrading the symmetry class of localization from unitary to symplectic class. 

Effects of disorder on topological insulators have been extensively studied.
\cite{OAN, Obuse, IKN, RS, jiang09, TAI, TTAI, guo, prodan10, goswami}
Directly relevant to this Letter are refs. \cite{OAN, Obuse, IKN}, which have treated 
the 
$s_z$ non-conserving
cases of  $\mathbb{Z}_2$TI, and refs. \cite{TAI, jiang09, TTAI} dealing with TAI. 
The basic viewpoint on TAI on which we are based is due to ref. \cite{TTAI},
i.e.,
spin-independent potential disorder brings about
renormalization of the gap, both its magnitude and sign,
allowing for conversion of an ordinary band insulator to a topologically non-trivial one.
On the other hand, 
a metallic behavior induced by disorder is also expected in the
% symplectic 
$s_z$ non-conserving
system we study.
Since the existing studies on TAI dealt with the 
% unitary 
$s_z$ conserving
case,
there was no discussion about
how such metallic conduction 
combines with the renormalization of band gap and localization effect
to determine the phase diagram of disordered BHZ model
with $s_z$ non-conserving SOC.
This paper takes full account of such Rashba-type spin-flip terms
incorporated in the BHZ model and deduces a phase diagram of
the disordered and disorder-induced $\mathbb{Z}_2$TI.
We demonstrate that
disordered BHZ model shows as many as four transitions 
for a given value of trivial band gap as a function of disorder.
As disorder increases, a narrow-gap topologically trivial insulator
first turns to a metal, second to a topological insulator, third again to a metal, 
and finally back to the trivial insulator.
Occurrence of such ``multiple transition" intervened by a finite metallic region
is a distinct property, providing us also with 
a useful point of view on the understanding of TAI.
% \textcolor{red}
% {Furthermore, 
% we show that existence of ``TAI" corresponds to enlargement of $\mathbb Z_2$TI in (band gap, disorder strength)-parameter space.}

We consider 
a tight-binding version of the BHZ model 
on 2D square lattice.  We add
 Rashba-type SOC in order to generalize to include the electric field perpendicular to surface, which is induced by structural inversion asymmetry of the quantum well or gate electrode.
Without disorder, the Hamiltonian becomes block diagonal
in momentum space and reduces to the following $4\times 4$
matrices,
\cite{BHZ, HgTe_JPSJ, BHZ+R, rothe10}
\begin{eqnarray}
{\mathcal H}
 (\bm k) &=
\left[
\begin{array}{cc}
h (\bm k)  &  \Gamma (\bm k) \\
\Gamma (\bm k)^\dagger & h^* (- \bm k)       
\end{array}
\right].
\label{ham_k}
\end{eqnarray}
The first (last) rows/columns correspond to real spin $\uparrow$ ($\downarrow$),
i.e., $s_z=\pm 1$.
The $2\times 2$ matrix $\Gamma(\bm k)$ represents Rashba SOC, the explicit form
of which will be determined by symmetry arguments.
The diagonal blocks, $h(\bm k)$ and its Kramers' partner
$h^* (-\bm k)$, are of 2D Dirac form: e.g.,
$h(\bm k) =  \bm d(\bm k) \cdot \bm \sigma$,
where
$$
\bm d(\bm k)= (A \sin k_x, A \sin k_y, \Delta-2B(2-\cos k_x -\cos k_y)).
$$
$\bm\sigma$ is another set of Pauli matrices (than the real spin $\bm s =(s_x, s_y, s_z)$),
representing an orbital pseudo-spin.
The parameter $\Delta$, which is associated with the band gap,
can be tuned by varying the thickness of HgTe layer.
Then nontrivial phase with $\mathbb{Z}_2$ topological invariant $\nu =1$
realizes when $0<\Delta /B <4$ and $4<\Delta /B <8$,
whereas trivial phase with $\nu =0$
corresponds to $\Delta /B <0$ and $\Delta/B > 8$.
% As for the off-diagonal blocks of Eq. (\ref{ham_k}),
% we keep only the diagonal terms of $\Gamma(\bm k)$:

In the presence of a random potential,
the total Hamiltonian is defined only in real space,
\begin{align}
H = \sum_{\bm r}  \left[
c^\dag_{\bm r} \epsilon_{\bm r} c_{\bm r}+
\left(
c^\dag_{\bm r+{\bm a}} t_x c_{{\bm r}}+
c^\dag_{{\bm r}+{\bm b}} t_y c_{{\bm r}}+
h.c.
\right)
\right],
\label{tb}
\end{align}
where
${\bm r}=(I,J)$ represents a site on 2D square lattice, and
${\bm a} = (1, 0)$ and ${\bm b} = (0, 1)$ are primitive lattice vectors 
with the lattice constant set to unity.
Explicit form of the hopping matrices
$t_x$ and $t_y$ are determined as follows.
We take the Rashba-type interaction,
$H_{\rm R} \propto (\bm p \times \bm s)_z = p_x s_y - p_y s_x$,
and adapt it to the BHZ model.
% for the $s_z$ non-conserving SOC. 
The two components of the orbital pseudo-spin
consist of $\Gamma_6$ orbital with the total angular momentum $j=1/2$,
and one of $\Gamma_8$ orbitals with $j=3/2$ and $j_z=\pm 3/2$.
% Only a limited number of matrix elements of $H_{\rm R}$
The form of $H_{\rm R}$ is constrained
due to various symmetry requirements.
First, notice that hopping matrix elements transform 
under $\pi$ rotation with respect to the $y$-axis
as,
\begin{align}
\alpha/2 & \equiv \langle \bm a, j_z | H_{\rm R} | \bm 0, -j_z \rangle
 = -\langle -\bm a, -j_z | (-H_{\rm R}) | \bm 0, j_z \rangle \nonumber\\
& = \langle \bm 0, j_z| H_{\rm R} | -\bm a, -j_z  \rangle^*.
\label{pi-rotation}
\end{align}
Here, particle-hole symmetry is assumed, implying that
the matrix elements considered are independent of $j_z$.
Combined with translational invariance, eq. (\ref{pi-rotation})
guarantees that $\alpha$ is {\it real}.
Second,
invariance under $\pi/2$ rotation with respect to the $z$-axis
requires, on the other hand,
\begin{align}
\langle \bm a, j_z | H_{\rm R} | \bm 0, -j_z \rangle
% = e^{i2j_z}  = \exp (i  j_z \pi)
& =\pm i
 \langle \bm b, j_z | H_{\rm R} | \bm 0, -j_z \rangle \nonumber \\
& = - \langle -\bm a, j_z | H_{\rm R} | \bm 0, -j_z \rangle,
\label{pi/2} 
\end{align}
where a positive (negative) sign corresponds to $j_z=\pm 1/2$ 
($\pm 3/2$).
Note that $\pi/2$ rotation gives rise to a factor, 
$\exp (i  j_z \pi)=\pm i$.
%\textcolor{red}{
%}
% , as required by $\pi$ rotation about the $y$-axis.}
% On the other hand,
% time-reversal symmetry requires the relation: 
%between right-top and left-bottom parts of hopping matrix as
% \begin{align}
% \langle \bm a, j_z |H_{\rm R}| \bm 0, -j_z \rangle = - \langle \bm a, -j_z | H_{\rm R} | \bm 0, j_z \rangle^*
% \equiv \alpha/2,
% \label{alpha} 
% \end{align}
% where the strength parameter $\alpha$ is real,
% as seen by comparing eqs.(\ref{pi/2}) and (\ref{alpha}).
By combining eqs. (\ref{pi-rotation}) and (\ref{pi/2}),
hopping matrices $t_x$ and $t_y$ are determined as,
\begin{align}
 t_x &= B \sigma_z - i \frac{A}{2} \sigma_x s_z + i \frac{\alpha}{2} s_y,
 \\
 t_y &= B \sigma_z + i \frac{A}{2} \sigma_y - i \frac{\alpha}{2} \sigma_z s_x.
\end{align}
Thus off-diagonal coupling in eq. (1) has been determined as,
\begin{align}
\Gamma(\bm k) = i \alpha \,
{\rm diag} \left[\sin k_x - i \sin k_y, \
\sin k_x + i \sin k_y \right].
\label{BHZ+R}
\end{align}
Terms connecting $j_z= \pm 1/2$ and $j_z = \mp 3/2$ states are neglected,
for simplicity.
% that respects four-fold rotational and particle-hole symmetries.
% The hopping terms are expressed in terms of the parameters, $A$, $B$ and $\alpha$ as,
% \begin{eqnarray}
% t_x = B\sigma_z -i \frac{A}{2} \sigma_x s_z + i \frac{\alpha}{2} s_x,
% \nonumber \\
% t_y = B\sigma_z +i \frac{A}{2} \sigma_y + i \frac{\alpha}{2} \sigma_z s_y.
% \label{hopping}
% \end{eqnarray}

For the impurity potential, we take a site-diagonal form.
The local energy $\epsilon_{\bm r}$ at site $\bm r$ is given by
\begin{align}
\epsilon_{\bm r} = (\Delta-4B) \sigma_z \otimes\mathbb{I} +
{\rm diag} \left[W_{\bm r}^{(s)}, W_{\bm r}^{(p)}\right]
\otimes\mathbb{I}
\end{align}
where 
$\mathbb{I}$ is $2\times 2$ identity matrix in spin space, 
and 
strength of non-magnetic 
disorder is specified by
$W^{(s)}_{{\bm r}}$ and $W^{(p)}_{{\bm r}}$ 
for $\Gamma_6$- and $\Gamma_8$-type orbitals,
respectively.
\cite{BHZ}
It is assumed that each impurity potential obeys 
uniform distribution in the period: $[-W/ 2, W/2]$.

% We briefly derive the form of $s_z$ non-conserving hopping.
% The hopping matrix is defined by
% \begin{align}
% t_\mu = \langle \bm n + \bm a_\mu, \Gamma',j_z' | H | \bm n, \Gamma, j_z \rangle.
% \end{align}

We have performed extensive numerical study
on the localization length $\lambda_L$ 
using the transfer matrix method.
\cite{TMM}
For a given set of parameters ($\Delta, W$),
the localization length $\lambda_L$ is estimated 
by varying the system size $L$.
If $\lambda_L /L$ decreases (increases)
with increasing $L$,
the system is judged to be insulating 
(metallic).
Hereafter, we concentrate on the case of $E=0$, where the Fermi level is located in the middle of band gap.
The other parameters are fixed as $A=B=1$.
We identify the localization characteristics from scaling behavior of the localization length $\lambda_L$, 
and determine the phase boundaries between metallic and insulating regions.

\begin{figure}
\includegraphics{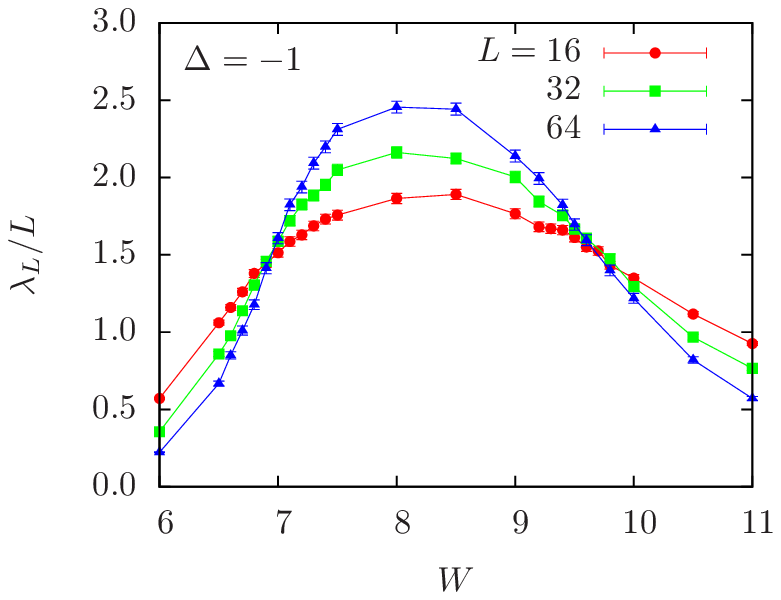}
\includegraphics{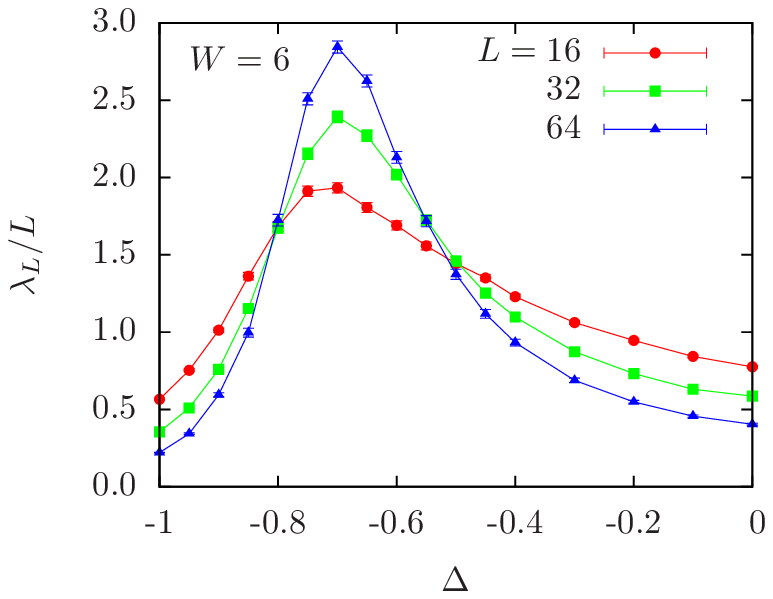}
\caption{(color online)
Scaling behavior of $\lambda_L$ as a function of $W$ (upper panel) and $\Delta$ (lower panel).
In the upper panel,  the system with $\Delta = -1$ 
is topologically trivial in the clean limit.
The lower panel shows the behavior
of "$\mathbb{Z}_2$ topological Anderson insulator" \cite{TAI, jiang09, TTAI},
since the region $\Delta > -0.5$ is made topologically nontrivial by the effect of $W$.}
\label{lambda}
\end{figure}

Figure \ref{lambda} shows representative data for different size 
of the system in the presence of $s_z$ non-conserving SOC with $\alpha=0.5$.
In the upper panel,  $\lambda_L /L$ is plotted as a function of
$W$  with $\Delta=-1$ fixed.  
At each value of $W$, $\lambda_L $ is evaluated for systems of different circumference $L$.
It is clear that there are two values of $W$ where
$\lambda_L /L$ does not depend on $L$.  This indicates a critical point that separates metallic and insulating regimes.
In contrast with ordinary phase transitions, the critical point is meaningful only at zero temperature, as an example of the quantum phase transition.
As $W$ increases from zero, the system becomes metallic
at $W=W_1 \sim 7$ up to $W_2 \sim 9.5$.
When $W$ is further increased, $\lambda_L /L$ shows again
a monotonic decrease as a function of $L$.
Namely we observe a {\it reentrant} behavior: 
localized $\rightarrow$ metallic $\rightarrow$ localized, with increasing $W$.
The emergence of metallic state by the $s_z$ non-conserving SOC has already been studied by ref.\citen{OAN} for the Kane-Mele model of graphene,
which belongs to a topological insulator in the clean limit.
In our case, however, 
the value $\Delta=-1$ means that the system is the trivial insulator in the clean limit.
Namely, in contradiction to naive intuition, the disorder first converts insulator to metal, and then back to insulator. 

%%%%%%%%%%%%%%%%%%%%%%%%%%%%%%%%%
%%%%%%%%%%%%%%%%%%%%%%%%%%%%%%%%%

The lower panel of Fig. \ref{lambda} shows
$\lambda_L /L$ as a function of $\Delta$ with fixed
$W=6$.
At each value of $\Delta$, $\lambda_L /L$ is evaluated for systems of different circumference $L$.
Overall behavior of 
$\lambda_L /L$ looks similar to that of the upper panel.
Namely,  as $\Delta$ increases, the system 
evolves as: 
localized $\rightarrow$ metallic $\rightarrow$ localized.
Note that 
the first insulating phase:
$\Delta < \Delta_1 \sim - 0.8$
is of trivial nature ($\nu=0$), whereas the second one with
$\Delta>\Delta_2 \sim - 0.5$) 
is $\mathbb{Z}_2$ nontrivial ($\nu=1$).
These different kinds are separated by
a finite metallic region: 
$\Delta_1 < \Delta < \Delta_2$.
The second insulating phase with $\mathbb{Z}_2$-number $\nu=1$ starts already
at a negative value: $\Delta_{c_2} \sim - 0.5$.
This result demonstrates that a 
$\mathbb{Z}_2$ topological insulator can be induced
by introducing non-magnetic disorder to a clean trivial insulator.
In recent literature, 
similar disorder-induced $\mathbb Z_2$-nontrivial phase has been discussed 
\cite{TAI, jiang09, TTAI},
but without the intervening metallic region.

By repeating such analysis for different values of $\Delta$ and $W$, and identifying critical points,
we obtain the phase diagram of the system at the ground state.
%%%%%%%%%%%%%%%%%%%%%%%%%%%%%%%%%
\begin{figure}
\begin{center}
\includegraphics{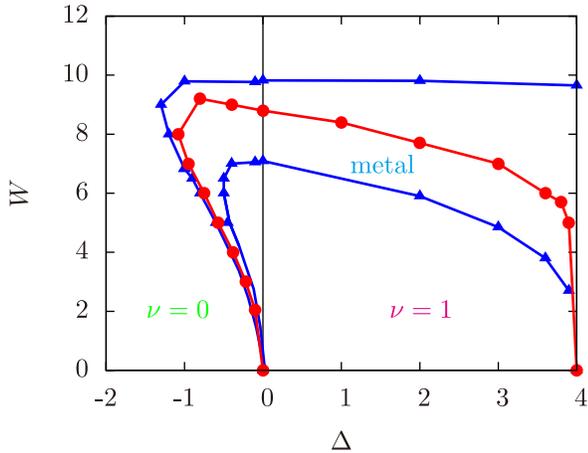}
\end{center}
\caption{(color online) Phase diagram of disordered BHZ model 
in the presence ($\alpha = 0.5$, blue triangles) and absence ($\alpha =0$, red circles) of  $s_z$ non-conserving SOC.
%Main part of the metallic phase, partitioning the two topologically distinct phases, appears as a broad band located in the range of disorder strength $W \sim 6-10$. It is extended toward the origin of ($\Delta$, $W$)-plane by a narrow corridor via the $\mathbb{Z}_2$-trivial side: $\Delta <0$.
Lines connecting the data points are guide to eyes.
}
\label{d_phase}
\end{figure}
%%%%%%%%%%%%%%%%%%%%%%%%%%%%%%%%%
%
In the clean limit, the nontrivial phase appears in the region 
$0<\Delta<4$ and $4<\Delta<8$.
Only the region with $\Delta <4$ is shown in Fig.\ref{d_phase} since
the phase diagram is reflection symmetric about 
$\Delta=4$.
The particle-hole symmetry is responsible for this property.
For comparison, we also show the
$s_z$ conserving case with $\alpha=0$.
In this case the system is always insulating except along the transition line indicated by (red) circles.

Let us focus on 
the $s_z$ non-conserving ($\alpha \ne 0$) SOC effects.
The triangle symbols (blue) show the transition points between metallic and insulating phases.
It is clear that the metallic region emerges in the vicinity of the transition line at $\alpha=0$.  
Consequently,
the two topologically distinct insulating phases are always separated by a metallic region of finite width.
Furthermore, 
$\nu =1$ phase is extended toward $\Delta <0$
by finite disorder.
%, which eventually destroys 
%the $\nu =1$ phase 
%at $W \sim 6$-$7$ if $\Delta \gtrsim -1$.
Unfortunately, in the weak-disordered region below $W \sim 4$ with $\alpha=0.5$, we have been unable to determine the transition point from the data up to 64 sites due to strong finite-size effect.
% \textcolor{red}{
The transition can occur, on the other hand, only at $\Delta=0$ in the clean limit, and the transition line should be continuously connected.
Therefore, with decreasing $W$, 
it is reasonable to assume that 
the metallic corridor persists and converges to the point $\Delta=0$.%}
As a result,
increasing the strength of disorder $W$,
multiple transition occurs in such a way
as: \\
(i) $\mathbb Z_2$-nontrivial $\to$ metallic $\to$ $\mathbb Z_2$-trivial for positive $\Delta$;  \\
(ii) $\mathbb Z_2$-trivial $\to$ metallic $\to$ $\mathbb Z_2$-nontrivial $\to$ metallic $\to$ $\mathbb Z_2$-trivial for $ -0.5 \lesssim \Delta < 0$.\\
These multiple transition are to be contrasted with the case $\alpha=0$ where 
intervening metallic phases are absent.

It is natural to ask the origin of the reentrant behavior with negative $\Delta$.
In order to answer the question,  we now turn to the density of states $\rho (E)$
of the system.
Let us first study how the region of nonzero $\rho (0)$ is correlated to the region of metallic conductance.
We follow the previous study \cite{TTAI} 
for the $s_z$ conserving case ($\alpha =0$), 
to employ the self-consistent Born approximation (SCBA).
Note that 
interference of electronic wave functions, which is crucial for the Anderson localization, is beyond the scope of the SCBA.
However, the SCBA does describe disorder-induced renormalization of $\Delta$, whose sign distinguishes whether the system is topologically trivial or not. 

Effects of disorder are taken into account in the SCBA as the self-energy $\Sigma (E)$ of the Green function
$G(\bm k, E)$, which is the $4\times 4$ matrix in our case.
We decompose the self-energy matrix as $\Sigma = \Sigma_0 + \Sigma_z \sigma_z$.
Note that the self energy is a scalar 
in the (real) spin 
%($\bm s$) 
space due to time-reversal symmetry. 
The renormalization of parameters occurs as
\begin{align}
\Delta & \rightarrow\tilde{\Delta}(E)=\Delta + {\rm Re}\ \Sigma_z(E), \\ 
E & \rightarrow\tilde{E}(E)=E - {\rm Re}\ \Sigma_0(E),
\end{align}
where $|\tilde{\Delta}|$ represents the renormalized energy gap, and $\tilde E$ is the shift of the energy.
The SCBA gives the following self-consistency equation:
\begin{align}
\Sigma (E) = \frac{W^2}{12} \int \frac{d^2k}{(2\pi)^2} \langle G(\bm k, E) \rangle,
\label{SCBA_def}
\end{align}
where 
$\langle G(\bm k, E) \rangle = \langle (E - {\mathcal H} (\bm k)-\Sigma (E))^{-1} \rangle$ 
is the disorder-averaged Green function.
Because of particle-hole symmetry, $\mathrm{Re} \ \Sigma_0(0)$ and $\mathrm{Im} \ \Sigma_z(0)$ vanish.
Hence the Fermi level at $E=0$ is not shifted.
The density of states is given by
\begin{align}
\rho (E) = - \frac{1}{\pi} \ {\rm Im} \ {\rm tr} \ G (E)
 = -\frac{24}{\pi} \ \mathrm{Im} \ \Sigma_0(E).
\end{align}
We solve Eq. (\ref{SCBA_def}) 
and derive
$\mathrm{Im} \ \Sigma_0(0)$ and $\mathrm{Re} \ \Sigma_z(0)$.
Near the metal-insulator transition, $\mathrm{Im} \ \Sigma_z(0)$ is small and 
converges only slowly in the iterative method.
Therefore we apply a root finding procedure called Steffensen's method only for $\mathrm{Im} \ \Sigma_0(0)$ in order to accelerate the convergence.

\begin{figure}
\includegraphics[scale=0.99]{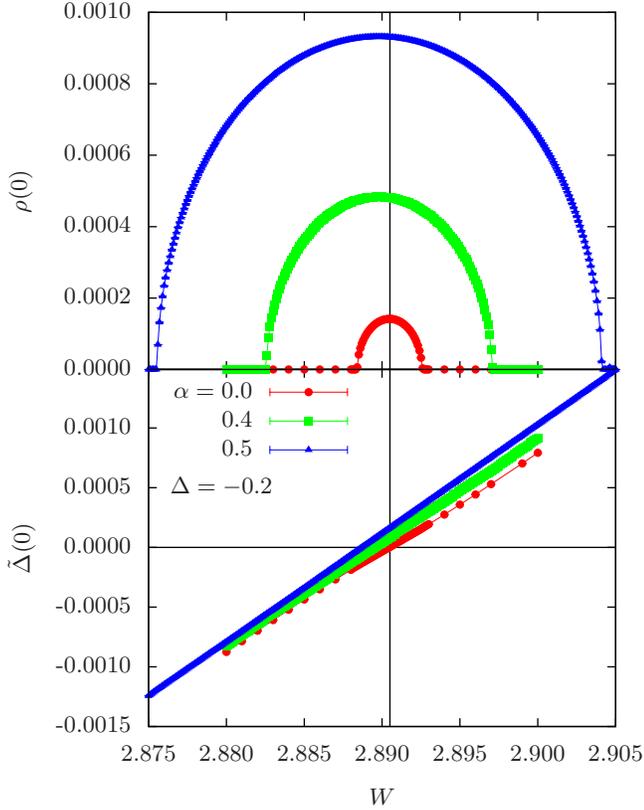}
\caption{(color online)
Density of states $\rho (0)$ (upper part) and renormalized mass $\tilde{\Delta}$ (lower part) 
%calculated in SCBA and plotted 
as a function of $W$ with fixed $\Delta = - 0.2$ for different values of $\alpha$. In the SCBA, $\tilde \Delta$'s for various $\alpha$ vanish commonly at $W \sim 2.89$, which corresponds roughly to the peak of $\rho (0)$.}
\label{dos}
\end{figure}

Figure \ref{dos} shows the density of states $\rho (0)$ (upper part)
together with the renormalized gap $\tilde \Delta(0) = \Delta + \Sigma_z(0)$ (lower part)
at the particle-hole symmetric point $E=0$.
%%%%%%%%%%%%%%%%%%%%%%%%%%%%%%%%
%%%%%%%%%%%%%%%%%%%%%%%%%%%%%%%%%
Results at fixed $\Delta = - 0.2$ but with different values of $\alpha$ is given 
for comparison.
In the upper part $\rho (0)$ is plotted as a function of $W$, 
which shows in all cases a finite range of finite $\rho (0)$.
According to the SCBA, the density of states $\rho (0)$ vanishes outside this range.
The change of $\rho (0)$ as a function of $W$ with decreasing $\alpha$ seems continuous down to the limit $\alpha=0$.
%, which means vanishing of $s_z$ non-conserving SOC.
Both the width and the magnitude of the region with
finite $\rho (0)$ decreases with decreasing $\alpha$, but 
remains finite at $\alpha =0$.

The lower part in Fig. \ref{dos} shows that 
$\tilde{\Delta}(0)$ changes sign around
$W \sim 2.89$, which corresponds roughly to the peak of $\rho (0)$.
Note that the bare value in the present case is $\Delta = - 0.2$, which corresponds to the trivial insulator with $\nu=0$.
Thus we find that the disorder drives the system into topologically nontrivial regime by renormalizing $\tilde{\Delta}(0)$ to a positive value.
The change of $\tilde{\Delta}(0)$ depends 
only slightly on $\alpha$ as shown in Fig. \ref{dos}.
Correspondingly,
%$\Delta\sim -0.2$, which is weakly on the trivial side, the density of states 
$\rho (0)$ becomes finite in a finite range of $W$ on both sides of the band inversion point ($\tilde{\Delta}=0$).
%%%%%%%%%%%%%%%%%%%%%%%%%%%%%%%%%%%%%%%%%%%%%%%%
\begin{figure}
\includegraphics{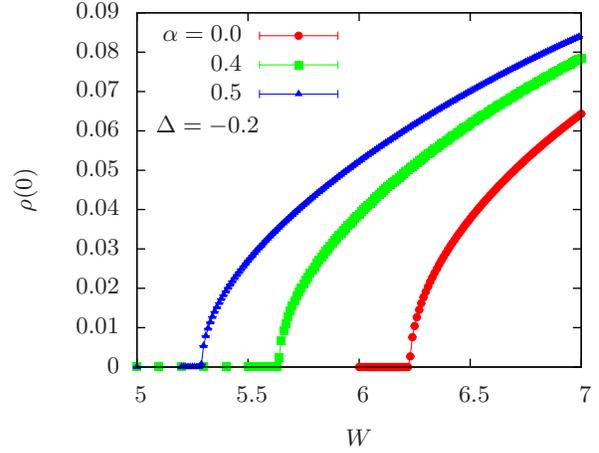}
\caption{(color line)
Density of states in the strongly disordered region for the same parameters as in Fig. \ref{dos}.}
\label{dos3}
\end{figure}
%%%%%%%%%%%%%%%%%%%%%%%%%%%%%%%%%%%%%%%%%%%%%%%%

As $W$ is further increased, $\rho (0)$ vanishes and 
stays null until the second onset of the topological transition as shown in Fig. \ref{dos3}.
Finite value for $\rho (0)$ resumes slightly before the entrance to the metallic phase shown in Fig.\ref{d_phase}.
% Note that the range of $W$ in Fig. \ref{dos} is too small to show this reentrance.
We have found that,
after the second onset, $\rho (0)$ never decreases but continues to grow with increasing $W$.
In other words, the end of the metallic phase around $W=10$ shown in Fig.\ref{d_phase} cannot be reproduced by the SCBA.
Although the density of states is finite for large $W$, 
the wave function is actually Anderson-localized.
The SCBA cannot describe this localization property.
Hence, it will safely be said that 
the reentrant behavior of the phase transition is partly understood 
by the renormalization of $\tilde{\Delta}(0)$ and the onset of $\rho(0)$ within the SCBA, especially in the range of small $W$.

%On the other hand,
%around $\Delta\sim -0.45$, the two regions of finite $\rho (0)$ as a function of $W$ tend to merge.  
%Namely, the narrow corridor with smaller $W$
%is absorbed in the wider part centered around $W\sim 8$
%with finite $\alpha \sim 0.1-0.2$.
%The multiple reentrance behavior
%%$\mathbb Z_2$-trivial $\to$ metallic $\to$ $\mathbb Z_2$-nontrivial $\to$ metallic $\to$ $\mathbb Z_2$-trivial 
%at small and negative $\Delta$ with increasing $W$ in Fig. \ref{d_phase} 
%seems to come from 
%interplay of three effects: 
%(i) band inversion due to renormalization of $\Delta$, 
%(ii) appearance of sizable $\rho (0)$, 
%% due to disorder induced intermediate states, 
%and (iii) interference of electronic states scattered by impurities, i.e., Anderson localization.
%The SCBA captures the effects (i) and (ii), while the insulating behavior in the $\mathbb Z_2$ nontrivial phase is the interplay of (i) and (iii).

In conclusion, we have derived the phase diagram of disordered BHZ model in the presence of $s_z$ non-conserving SOC. 
%From the scaling behavior of the wave function studied by the transfer matrix method, we identify critical points in the parameter space of the band-gap parameter $\Delta$ and the disorder strength $W$. We have found multiple transitions from $\mathbb{Z}_2$ topologically trivial to nontrivial insulating phases. 
As a crucial effect of $s_z$ non-conserving SOC,  we have demonstrated that metallic state intrudes each transition.
While
TAI is realized as a distinct state in $(W,E)$-space\cite{TAI},
it is connected continuously to clean topological insulator in $(\Delta, W)$-space.
Furthermore, the origin of reentrant behavior is investigated in terms of the SCBA.
As in the case of $\alpha=0$ studied earlier \cite{TTAI},
we have found for $\alpha\neq 0$ 
that the disorder drives originally negative $\Delta$ to positive. 
This change of sign causes the topological transition induced by disorder.
However, as an important difference from the previous study, 
we find that metallic state always accompanies the topological transitions
in the phase diagram.
Actual 2D 
$\mathbb{Z}_2$TI systems such as HgTe quantum wells have a sizable $s_z$ non-conserving SOC, which can be tuned continuously by gate voltage.
We hope that our results encourage experimental detection of (multiple) topological transitions with metallic intruders.

AY acknowledges Grant-in-Aid for JSPS Fellows under Grants No. 08J56061.
KN and KI are supported by Grant-in-Aid for Young Scientists (B) under Grants 
Nos. 20740167 (KN) and 19740189 (KI).
KN is also supported by FIRST program (JSPS).
We acknowledge C. Br\"une and E. Prodan for 
helpful discussions and correspondences.

\bibliographystyle{jpsj}
\bibliography{DIMT}

\clearpage

\end{document}